\documentclass[conference]{IEEEtran}
\usepackage{cite}
\usepackage{amsmath,amssymb,amsfonts}
\usepackage{algorithmic}
\usepackage{graphicx}
\usepackage{tabularx}
\usepackage{booktabs}
\usepackage{textcomp}
\usepackage{xcolor}
\def\BibTeX{{\rm B\kern-.05em{\sc i\kern-.025em b}\kern-.08em
    T\kern-.1667em\lower.7ex\hbox{E}\kern-.125emX}}
\usepackage[bottom]{footmisc}
\usepackage{rotating}
\usepackage{amsmath}
\usepackage{amsthm}
\usepackage{amsfonts}
\usepackage{amssymb}
\usepackage{subfig}
\usepackage[sort]{natbib}
\usepackage{hyperref}
\setcitestyle{round} 

\graphicspath{/Figures}
    
\begin{document}

\title{Representation Learning for Regime Detection in Block Hierarchical Financial Markets}

\author{\IEEEauthorblockN{Alexa Orton} 
\IEEEauthorblockA{Department of Statistical Sciences\\ 
University of Cape Town\\ 
\texttt{aorton@marketaxess.com}\\
ORCID: 0000-0002-4698-0660 \\ \vspace*{-1cm}}
\and
\IEEEauthorblockN{Tim Gebbie}
\IEEEauthorblockA{Department of Statistical Sciences\\ 
University of Cape Town\\ 
\texttt{tim.gebbie@uct.ac.za}\\
ORCID: 0000-0002-4061-2621}}
\maketitle

\begin{abstract}
We consider financial market regime detection from the perspective of deep representation learning of the causal information geometry underpinning traded asset systems using a hierarchical correlation structure to characterise market evolution. We assess the robustness of three toy models: SPDNet, SPD-NetBN and U-SPDNet whose architectures respect the underlying Riemannian manifold of input block hierarchical SPD correlation matrices. Market phase detection for each model is carried out using three data configurations: randomised JSE Top 60 data, synthetically-generated block hierarchical SPD matrices and block-resampled chronology-preserving JSE Top 60 data \footnote{See block-resampled chronology-preserving data set evaluation at: \citet{AOrtonMSc2024}}. We show that using a singular performance metric is misleading in our financial market investment use cases where deep learning models overfit in learning spatio-temporal correlation dynamics.
\end{abstract}


%
\IEEEpeerreviewmaketitle

\section{Introduction}

Recent financial crises have made it ever-more important to detect shifts in macroeconomic regimes as underpinned by the hierarchical latent features and non-linear causal relationships driving market evolution \citep{miori2022returnsdriven,PapenbrockJochen2021MESC}. Novel approaches have been developed to address the key problems related to the assumptions of Guassianity, stationarity, and the need to address robustness and reflexivity \citep{MartiGautier2016CFTS,PGF2023ssrn}.

Here we consider the SPDNet introduced by \citet{huang2016SPDNet} to establish a benchmark for deep representation learning of the SPD manifold of hierarchical correlation matrices for regime detection. We compare this model's performance to that of the SPDNetBN of \citet{brooks2019SPDNetBN} to determine whether regime classification accuracy improves as a function of the {\it batchnorm} respecting the Riemannian manifold upon which the SPD correlation matrices sit. Finally, the U-SPDNet \citep{WANGUNET2022} is considered to evaluate model capacity for latent block hierarchical feature extraction in conjunction with regime detection accuracy as a result of an augmented Riemannian autoencoder structure and increased information pass-through. 

The real-world comparative scenario of Mean-Variance Portfolio Optimisation is then presented in Figure \ref{fig:backtesting} to provide a use-case for using these models for regime-dependent portfolio construction. Backtesting on data adjusted for information leakage using purged and embargo reveals that U-SPDNet provides marginally larger returns than SPDNet, while both under-perform the equally-weighted benchmark portfolio \citep{depradoML}.

\section{Data}

\subsection{Empirical Data}

\begin{figure*}[t!]
    \centering
    \subfloat[Stressed correlations]{
       \includegraphics[width=0.31\textwidth]{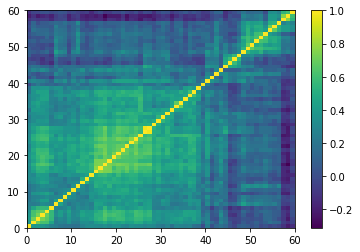}}
        \label{fig:sub1}
    \hfil
    \subfloat[Normal correlations]{\includegraphics[width=0.31\textwidth]{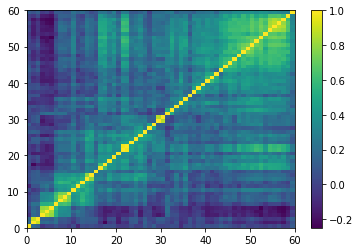}}
        \label{fig:sub2}
    \hfil
    \subfloat[Rally correlations]
    {\includegraphics[width=0.31\textwidth]{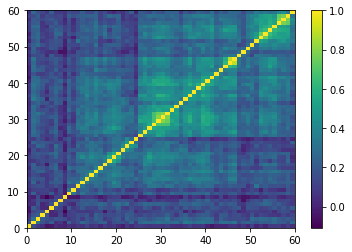}}
        \label{fig:sub3}
    \caption{Average correlation matrices for each regime identified from the JSE Top 60 data. Distance metric: $\sqrt{2(1 - C})$, where $C$ is the sample correlation matrix. Matrices are classified into three regimes according to the \emph{ex ante} Sharpe ratio ($SR$). Stressed: $SR < -0.5$; normal: $SR \in [-0.5,2.0]$; rally: $SR > 2.0$. In the stressed class we can see a large cluster in the lower left, as many assets have synchronised movements, as compared to the normal market condition where the block diagonal structure suggests improved diversification. Average rally regimes exhibit this same hierarchical structure with lower inter- and intra-block correlation strength.}
    \label{fig:corrmats}
\end{figure*}

\begin{figure}[h]
  \centering
  \includegraphics[width=0.98\linewidth]{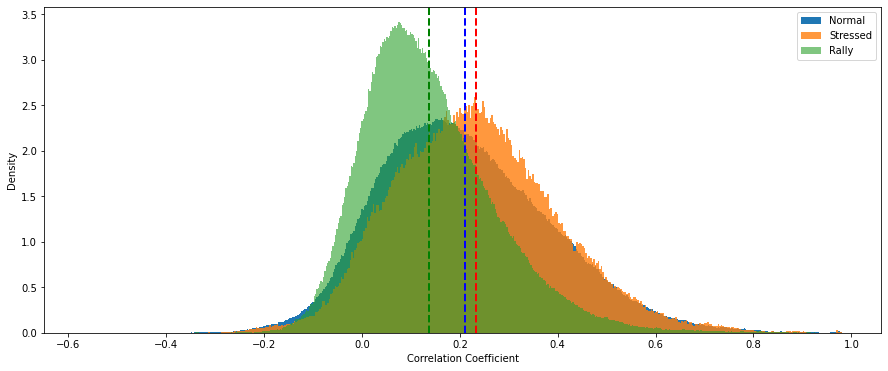}
  \caption{JSE Top 60 correlation coefficient density per class. The stressed, normal and rally regimes have mean correlations of 0.24, 0.205 and 0.17, respectively. Average correlation and standard deviation vary per regime, supporting the findings of \citet{miori2022returnsdriven} in the South African market of traded stocks, where varying underlying distributions characterise different macroeconomic regimes. Interrelatedness of assets is greater in times of market stress with greater volatility, while normal and rally regimes are characterised by lower average correlation and volatility. The regime correlation asymmetries identified by \citet{AngAndrew2004HRAA} prevail in the JSE top 60 data. The average correlation matrices for each market regime are shown in Figure \ref{fig:corrmats}.}
  \label{fig:corrcoeffdensity}
\end{figure}

To conduct a real-world application of the selected manifold learning techniques for regime classification, the set of daily equity returns of the top 60 high-capitalised stocks traded on the JSE over the period January 2000 - December 2023 is considered. JALSH price data are obtained via Bloomberg (BBG) which are then filtered according to a market capitalisation-determined ranking\footnote{See data processing code at: \citet{ortongebbie2024}}.

The JSE Top 60 stocks comprise an index of the most-traded stocks in the South African financial market. Considered through the lens of market regime detection, it is observed that the market passes through several significantly different structural periods characterised by particular price volatility dynamics. Specifically, stressed periods seen in 2007-2008 (Global Financial Crises), 2015-2016 and March 2020 (Covid-19) which were characterised by market downtrends and correspond with greater volatility periods in average price action. 

This conforms with the work of \citet{AngAndrew2004HRAA}, where bear markets are characterised by increased volatility. It is expected that the SPDNet, SPDNetBN and U-SPDNet models will be capable of detecting the differences between these stressed market states and their normal, or rally, counterparts which feature lower levels of price action volatility. We show that this is indeed the case.

\subsection{Synthetic Data}

\begin{figure*}[h!]
    \centering
    \subfloat[Stressed correlations]{
       \includegraphics[width=0.31\textwidth]{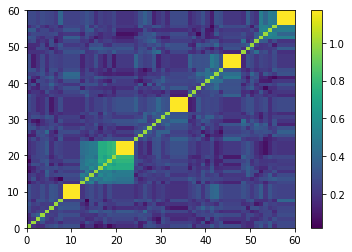}}
        \label{fig:sub11}
    \hfil
    \subfloat[Normal correlations]{\includegraphics[width=0.31\textwidth]{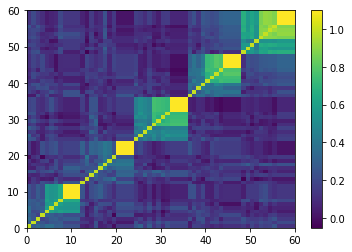}}
        \label{fig:sub22}
    \hfil
    \subfloat[Rally correlations]
    {\includegraphics[width=0.31\textwidth]{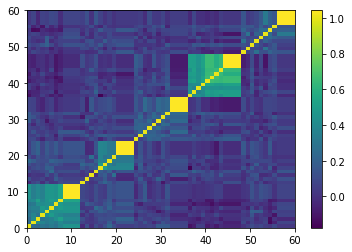}}
        \label{fig:sub6}
    \caption{We show representative (simulated) block hierarchical SPD correlation matrices per regime. The synthetic construction uses techniques for nested factor timeseries generation as proposed by \citet{yelibi2021ALC} to create 5 clusters with 3 hierarchies each, sampled from Student's t-distribution with $v = 3$ d.o.f. The  SPD correlation matrices computed on these datasets are permuted to introduce realistic noise with seed = 27. The simulated correlation matrices can be visually compared with the empirical average correlation matrices for each market regime as shown in Figure \ref{fig:corrmats}.}
    \label{fig:corrsynthmat}
\end{figure*}

\begin{figure}[h!]
  \centering
  \includegraphics[width=0.98\linewidth]{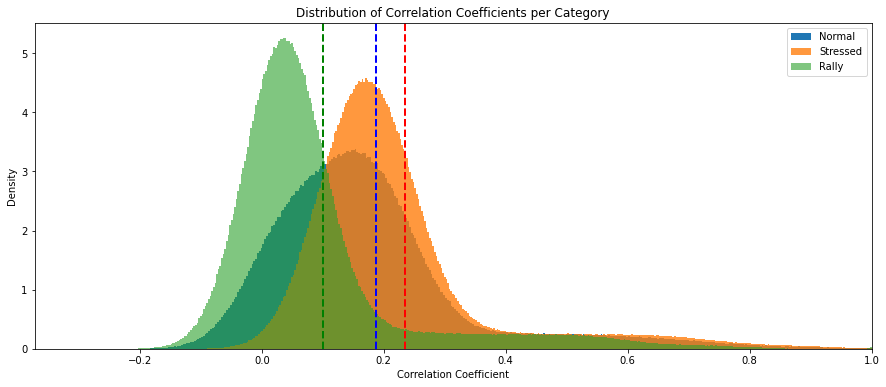}
  \caption{Synthetic block hierarchical SPD matrix correlation coefficient density per regime. By construction, the stressed, normal and rally regimes have average correlations of 0.24, 0.18 and 0.10 respectively, with descending standard deviation values. The separability of underlying distributions per regime class is preserved to emulate Figure \ref{fig:corrcoeffdensity}.}
  \label{fig:corrsynth}
\end{figure}

Synthetic correlation structures are generated using the method of \citet{yelibi2021ALC} where multiple factors are incorporated in a nested fashion into the price time series model, or into the benchmark HNFM originally proposed by \citet{tumminello2007hierarchically}. No time factor is incorporated in the generation of these timeseries or associated correlation matrices. These provide a set of counterfactual scenarios characterised only by a hierarchically nested structure, where no information lead-lag effects are present in determining the market regime.

Price returns processes are modelled with random variation in the $\beta$ factor sensitivity value in conjunction with additive perturbations in $\eta$ and $\epsilon$ values to construct 18,000 time series for 60 data features in order to mirror the real-world price action data set of BBG JSE top 60 equities for $T = 252$ per computation window \citep{yelibi2021ALC, marti2020}.

The correlation matrices computed on these synthetic time series preserve the SPD structure, and can then be used via a Cholesky factor decomposition-motivated Monte Carlo simulation to conduct the SR-guided market regime labelling \citep{PapenbrockJochen2021MESC}. In order to respect the empirically-determined and well-documented fat-tailed nature of financial asset returns time series, this simulation uses a Student-t distribution drawing from an $N$-variate distribution with $v = 3$ d.o.f such that $X \sim t_v \left(0, \frac{v-2}{v}\Sigma \right)$ \citep{MartiGautier2016CFTS}. Having generated these time series, the correlation matrices are allocated to their relevant stressed, normal or rally category in accordance with the method of \citet{marti2021} (as in Figure  \ref{fig:corrmats} above) using the contemporaneous stock  basket $SR$.

By construction, the synthetic SPD matrix correlation coefficients display distributions in line with the regime switching models introduced by \citet{AngAndrew2004HRAA} and identify different macroeconomic regimes being modeled via making draws from two or more possible underlying distributions. This set of distributions is fabricated in the interest of preserving the stylised facts of real-world regime correlation dynamics, and we report that this is evidenced by the closeness in market regime distributional patterns observed between the synthetic data and the empirical data. We note in particular that the synthetic matrices possess a hierarchical structure and their pairwise distributions are positively-shifted (See figures \ref{fig:corrcoeffdensity} and \ref{fig:corrsynth}).

\section{Market Regimes}

Market-timing strategies which exploit changing regime-specific underlying data distributions have been shown to add value to portfolio protection products {\it e.g.} TIPP and risk-off investment decisions; not only confined to those concerning equities \citep{AngAndrew2004HRAA,PapenbrockJochen2021MESC}.

\citet{AngAndrew2004HRAA} identify two broad market regimes, namely normal and bear markets. Using regime switching (RS) asset returns model for portfolio asset allocation, the regimes are accurately reconstructed by assuming returns data points may be drawn from differing underlying data generation distributions, which supports the empirically-observed asymmetric correlation distribution. Recently, centrality measures computed on correlation coefficients representing evolving equity financial market networks have been employed in matrix clustering, regime detection and risk-diversified portfolio construction by extension \citep{BorghesiChristian2007Eoti}. 

\citet{miori2022returnsdriven} extend these studies to returns correlations across multiple asset classes and establish a purely data-driven technique for distinguishing regimes. Influenced by spectral analysis techniques, cophenetic correlation matrices are constructed as correlations between the distance matrix found as $\sqrt{2(1-\|C\|)}$ where $C$ is Pearson's linear correlation, sitting on the Euclidean manifold, and the associated dendrogram generated by a hierarchical linkage algorithm. Six regimes are uncovered via KMeans++ clustering using the elbow method for cluster number selection. 

Here three classes of correlation matrices are generated by augmenting the original data which correspond to stressed, normal and rally market regimes. This marries the findings of \citet{miori2022returnsdriven} and \citet{AngAndrew2004HRAA}, and emulates the 'rule-of-thumb' technique used by \citet{marti2021}. By virtue of considering that three broad market regime categories have been detected in equity markets, and that the annualised $SR$ of indices belonging to multiple asset classes differs significantly amongst six data-informed regimes, a categorisation system emerges for correlation matrix labelling.

\begin{figure}[!h]
\centering
\subfloat[JSE Top 60]{\includegraphics[width=0.5\textwidth]{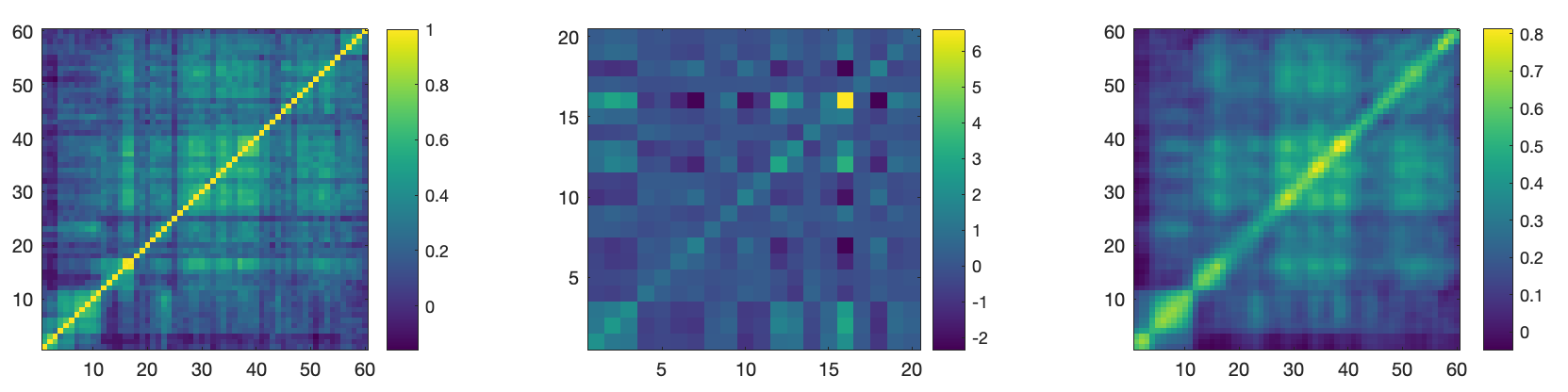}}
\label{fig:latent60}
\hfil
\subfloat[Synthetic Data]{\includegraphics[width=0.5\textwidth]{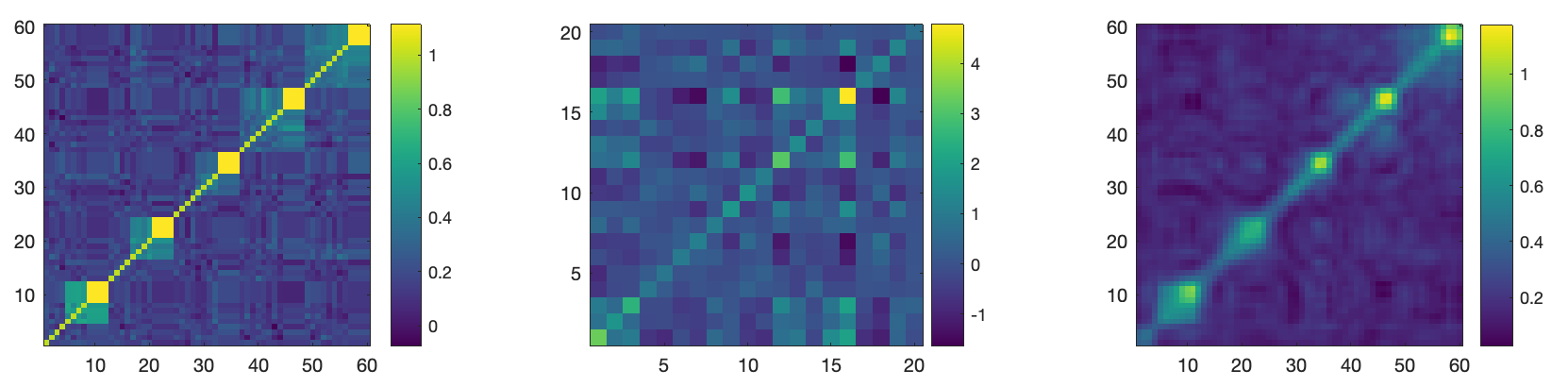}}
\label{fig:synth600}
\caption{U-SPDNet matrix evolutions. Here, the 60x60 input space (left), 20x20 latent space (center) and the reconstructed 60x60 layer (right) matrices of U-SPDNet trained and validated on both synthetic nested block hierarchical correlation matrices (lower row) and their JSE Top 60 data counterpart (upper row). Representational example drawn from the 600th epoch.}
\label{fig:matevols}
\end{figure}

\section{Learning Accuracy}

We evaluate the SPDNet-based models using both the ordered JSE top 60 correlation matrices and their block hierarchical synthetic data counterpart. Much is revealed by these initial experiments which provide a baseline specifically for market regime detection given the information contained in these datasets alone. 

Firstly, in comparing the 2-layer {60, 20, 3} and 4-layer {60, 40, 20, 10, 3} configurations, it is clear that the introduction of Riemannian \emph{batchnorm} layers improves the market state classification accuracy, irrespective of the dataset used for training, which confirms the findings of \citet{brooks2019SPDNetBN} where the integration of Riemannian batchnorm layers with the original SPDNet yielded greater radar image classification accuracy. 

Furthermore, as identified by \citet{WANGUNET2022}, increased depth holds no relationship with improved classification accuracy since the performance measure convergence occurs within a narrow band around 71\% and 66\% for the real-world and synthetic data scenarios, respectively, irrespective of model depth. Model performance is weaker when the synthetic block hierarchical correlation matrices are used as training input when compared to average validation accuracies achieved for their real-world counterparts.

All SPDNet benchmark models converging on training accuracy values nearing 70\% which resemble, and even improve on, those achieved when the self-same models are applied in the context of allocating correlation matrices computed from S\&P 500 stock price returns to the stressed, normal and rally categories - as in the work of \citet{marti2021}.

\begin{table}[h!]
  \centering
  \caption{Model Configurations. Model acronyms are shown in the first column. The table then provides the Transformation Matrix Dimension (TMD) that defines the architecture of the model learning systems, and the learning rate $\lambda$. The Momentum parameter $\eta$ is set to 0.9 for all models. The U-SPDNET-6BiRe model configuration is annealed over a range of learning rate parameters, and its TMD is mirrored across the architecture (**) where the Cross-Entropy (CE) loss function is guided by the Reconstruction Error Term (RET). BN indicates the use of Riemannian Batch Normalisation (RBN); BiRe the implementation of alternating BiMap and ReEig layers. The simulation accuracy results for the configurations are summarised in Table \ref{tab:accuracies}.}
  \label{tab:table1}
  \small
  \begin{tabular}{lllll}
    \toprule
    \textbf{Model} & \textbf{TMD} & \textbf{$\lambda$} \\
    \midrule
    SPDNet & \{60, 20, 3\} & $1 \times e^{-3}$  \\
    SPDNetBN & \{60, 20, 3\} & $1\times e^{-3}$ \\
    SPDNet-3BiRe & \{60, 40, 20, 10, 3\} & $1 \times e^{-4}$\\
    SPDNetBN-3BiRe &  \{60, 40, 20, 10, 3\} & $1 \times e^{-4}$  \\
    U-SPDNet-6BiRe & \{60, 40, 20, 10, 3\} (**) &  $1 \times [e^{-2}, e^{-5}]$ \\
    \bottomrule
  \end{tabular}
\end{table}

\begin{table}[h!]
  \centering
  \caption{Regime Detection Accuracy Out-Of-Sample (OOS). U-SPDNet outperforms all SPDNet benchmark models in the context of the empirical real-world data (JSE Top 60 data). The U-SPDNet OOS accuracy when trained on the synthetic dataset does not improve on the initial SPDNet model configurations. The various TMDs are shown with mirroring noted as above (**). This may suggest that the richer non-linear structure found in the real-world data is better leveraged by more complex architectures.}
  \label{tab:accuracies}
  \small
  \begin{tabular}{lllll}
    \toprule
    \textbf{Model} & \textbf{Configs.}&\textbf{Emp.} & \textbf{Synth.} \\
    \midrule
    SPDNet & \{60, 20, 3\} & 68.95\% & 66.80\% \\
    SPDNetBN & \{60, 20, 3\} & 72.46\% & 65.92\% \\
    SPDNet-3BiRe & \{60, 40, 20, 10, 3\} & 69.50\% & 64.94\% \\
    SPDNetBN-3BiRe & \{60, 40, 20, 10, 3\} & 70.71\% & 66.20\% \\
    U-SPDNet-6BiRe & \{60, 40, 20, 10, 3\} (**) & 98.90\% & 65.11\% \\
    \bottomrule
  \end{tabular}
\end{table}

\begin{figure}[h!]
  \centering
  \includegraphics[width=0.98\linewidth]{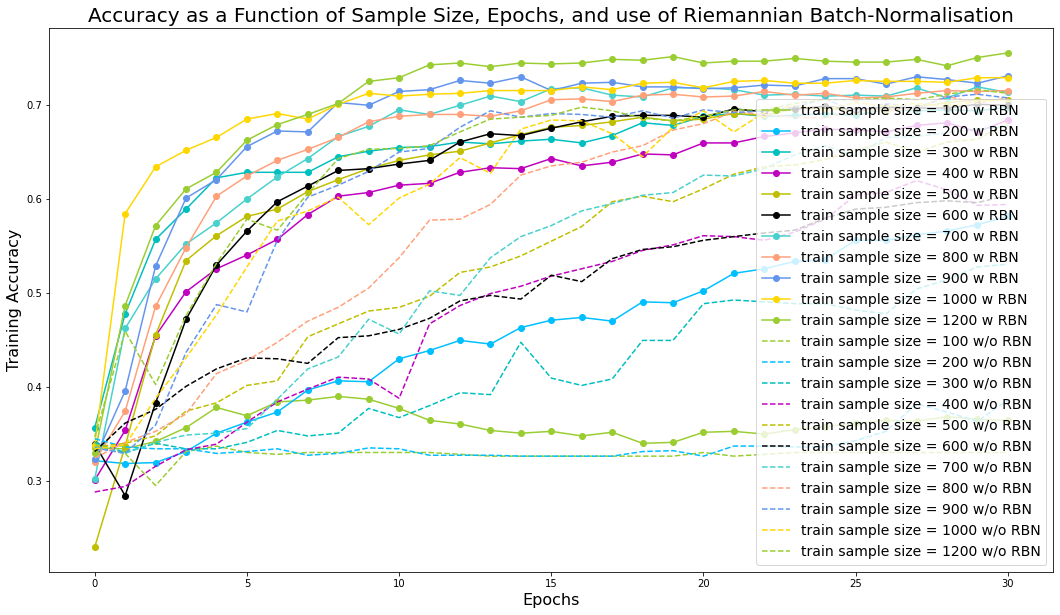}
  \caption{Accuracy of SPDNet trained on hierarchically-ordered JSE Top 60 correlation matrices. Distance metric: $\sqrt{2(1 - \mathbf{C}})$ where $\mathbf{C}$ is the correlation matrix. Training performed both with (w) and without (w/o) Riemannian Batch Normalisation (RBN). In both instances, training accuracy is greater for larger training sample sizes. We note that the introduction of RBN layers improves the market state classification, and that this effect is enhanced for larger sample sizes. }
  \label{fig:fig1acc}
\end{figure}

\begin{figure}[h!]
  \centering
  \includegraphics[width=0.98\linewidth]{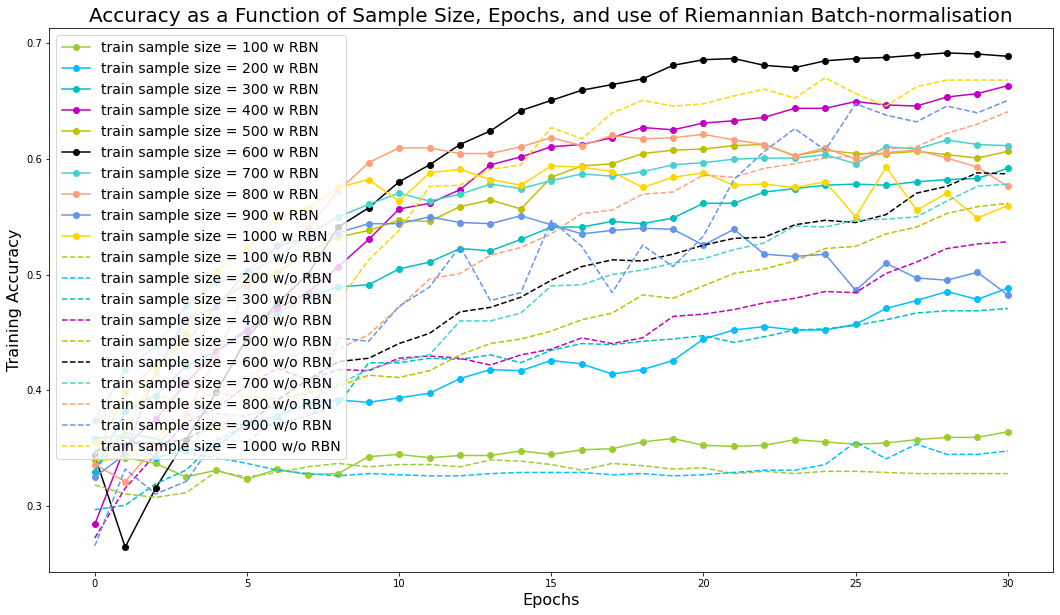}
  \caption{Accuracy of SPDNet-3BiRe trained on hierarchically-ordered JSE Top 60 correlation matrices. Distance metric: $\sqrt{2(1 - \mathbf{C}})$ where $\mathbf{C}$ is the correlation matrix. Training performed both with (w) and without (w/o) RBN. Similarly as found for SPDNet, we note that the introduction of RBN layers improves the market state classification. This can be compared to the SPDNet-3BiRe trained on the simulated dataset and suggests that training on empirical data is more effective, perhaps because of the enhanced existence of non-linear patterns in real world data.}
  \label{fig:fig2acc}
\end{figure}

\begin{figure}[h!]
  \centering
  \includegraphics[width=0.98\linewidth]{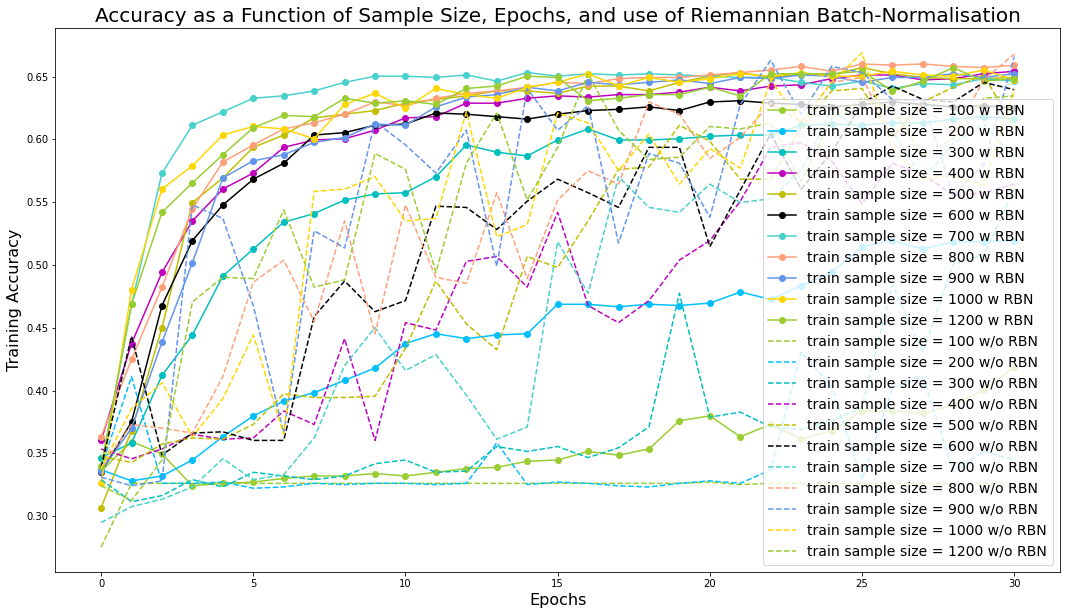}
  \caption{Accuracy of SPDNet trained on synthetic block hierarchical SPD correlation matrices. Training performed both with (w) and without (w/o) RBN. This may be compared to the SPDNet performance in Figure \ref{fig:fig1acc} using the empirical dataset, and we again note that the introduction of RBN layers improves the market state classification.}
  \label{fig:fig3acc}
\end{figure}

\begin{figure}[h!]
  \centering
  \includegraphics[width=0.98\linewidth]{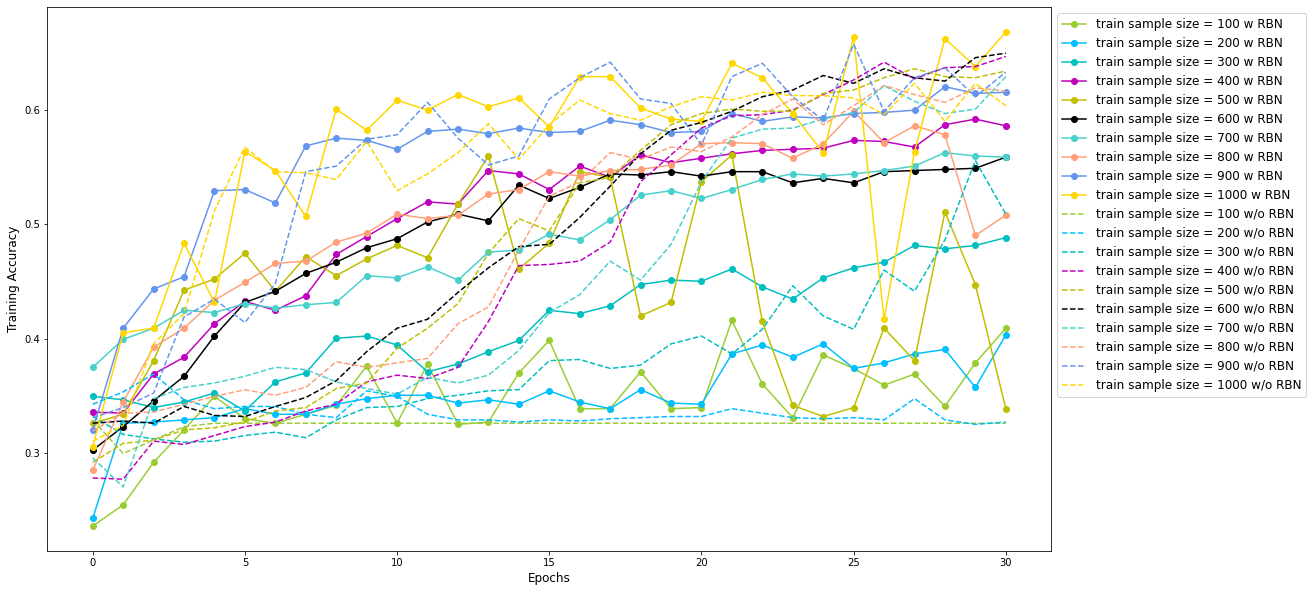}
  \caption{Accuracy of SPDNet-3BiRe trained on synthetic block hierarchical SPD correlation matrices. Training performed both with (w) and without (w/o) Riemannian Batch-Normalisation (RBN). This can be compared to SPDNet-3BiRe trained on empirical correlation matrices in Figure \ref{fig:fig2acc}.  The introduction of RBN layers can improve the market state classification, but the effect seems moderate as compared to that seen in SPDNet empirically-trained models.}
  \label{fig:fig4acc}
\end{figure}

\section{Results}

Figure \ref{fig:matevols} shows how the block hierarchical structure is maintained in the latent layer in both cases. In the JSE Top 60 data case, the reconstructed matrix preserves the block hierarchical structure, but a one-to-one information mapping is not perfectly achieved OOS. The extracted latent 20-feature layers show preservation of the block-hierarchical structure and meaningful statistical factor identification. The latent space matrices have similar structures with subtle differences, pointing to the U-SPDNet model as capturing a tractable underlying correlation information set on the SPD manifold which incorporates the temporal feature of the real world data in comparison to their synthetic counterparts.

\begin{figure}[h!]
    \subfloat[SPDNet confusion matrix]
    {\includegraphics[width=0.24\textwidth]{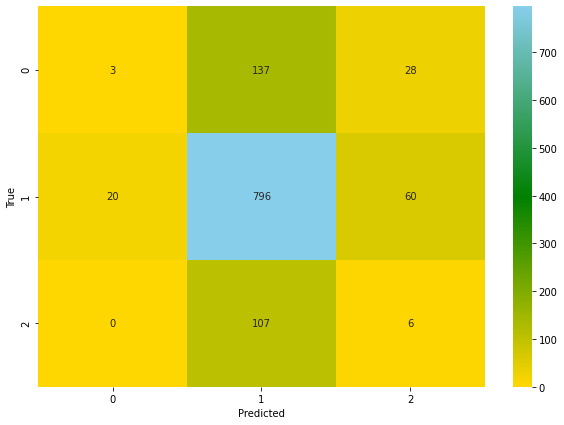}}
        \label{fig:sub31}
    \subfloat[U-SPDNet confusion matrix]
    {\includegraphics[width=0.24\textwidth]{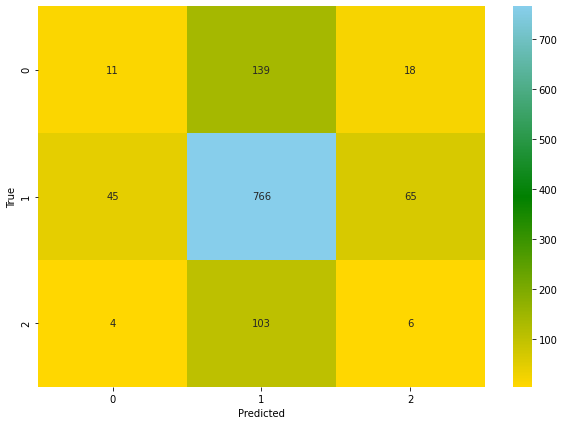}}
        \label{fig:sub32}
\caption{The confusion matrices depicts the model’s estimated regime labels, as compared to the SR-labeled correlation matrices computed on the test data set of JSE Top 60 returns on [01/2008 - 07/2012]. The SPDNet model makes most accurate predictions of normal regime labels, and inaccurately classifies stressed and rally labels as normal regimes in the majority of cases. Relative to the SPDNet-identified regimes, the stressed and rally regimes show improved predictive accuracy at the expense of marginally lower accuracy for normal regime detection when comparing the model’s predictions to the true $SR$-determined labels OOS.}
\label{fig:confusion}
\end{figure}

The confusion matrices in Figure \ref{fig:confusion} show that the superior prediction accuracy of the SPDNet OOS is a product of simply converging on a corner solution. This, despite generating many additional training samples to mitigate class imbalance, results in inaccurate predictions; the SPDNet fails in learning a representation of the hierarchical structure underpinning market evolution in different regimes as evidenced by it learning to place most test observations into the normal market regime. Thus, despite achieving lower regime detection accuracy,the U-SPDNet is more likely to have learned an improved information-geometric representation of the market’s hierarchical structure, which leads to marginally-greater state detection accuracy in stressed and rally classes.

\section{Conclusion}[h!]

U-SPDNet performs improved latent feature extraction with better classification performance in stressed and rally market phases, but achieves lower OOS backtest scenario accuracy than that of the benchmark SPDNet. The SPDNet-based models fail in capturing the latent spatio-temporal block hierarchical correlation dynamics and deliver corner solutions across all input data sets. However, U-SPDNet is promising in terms of its utility in regime-dependent portfolio optimisation strategy generation as a model better-suited to capturing latent block hierarchical correlation structures arising from lead-lag causal feedback information loops that often drive the evolution of market regimes. Here we highlight the importance of avoiding reliance on accuracy as a robust model performance measure in the context of complex financial markets data.

\begin{figure}[h!]
\centering
\includegraphics[width=0.47\textwidth]{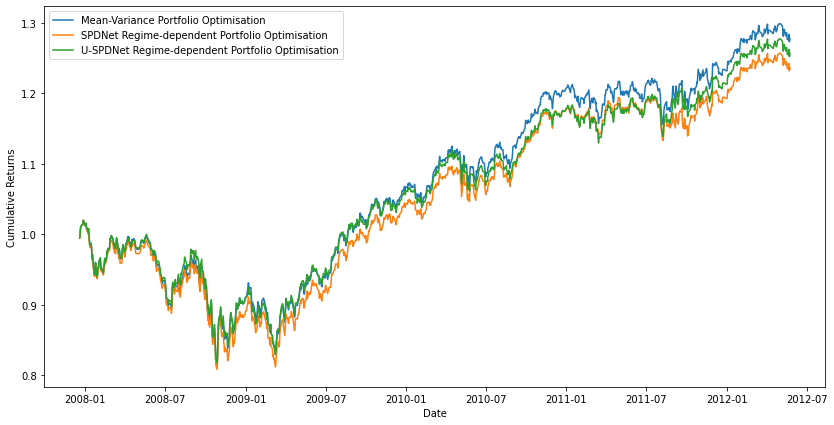}
\caption{MVPO backtesting. We perform daily-rebalanced portfolio optimisation backtesting on the OOS JSE Top 60 data set from 01-2008 - 07-2012 with no trading costs. Before 2010 the U-SPDNet regime-dependent strategy outperforms both the benchmark mean-variance and SPDNet regime-dependent strategies during the stressed market periods associated with the GFC. Thereafter, the U-SPDNet regime-dependent strategy outperforms its SPDNet counterpart; the mean-variance portfolio delivers greatest cumulative returns overall.}
\label{fig:backtesting}
\end{figure}

Figure \ref{fig:backtesting} shows the cumulative performance of the three portfolio optimisation strategies on the real-world JSE Top 60 dataset. Initially, the U-SPDNet delivers greatest returns; this regime-dependent daily rebalancing approach marginally outperforms the Mean-Variance portfolio until 2010. Thereafter, the Mean-Variance benchmark optimisation consistently outperforms both regime-dependent strategies. The SPDNet-informed regime-dependent optimisation performs worse particularly during market rally periods, which corresponds to this category being significantly under-detected by the model in the OOS period as shown in Figure \ref{fig:confusion}. 

\subsection{Research Reproducibility}
All code written in support of this research may be found at the \citet{ortongebbie2024} GitHub repository BH-SPD-CorrMat-Nets, or on the Deep Learning Block Hierarchical SPD Correlation Matrices for Regime Detection OSF project page \url{https://osf.io/msrgc/} where data for both SPDNet, SPDNetBH and U-SPDNet model evaluation on both real world and synthetic data is stored. 
Required repositories:
\begin{enumerate}
    \item \url{https://gitlab.lip6.fr/schwander/torchspdnet}
    \item \url{https://github.com/GitWR/U-SPDNet}
    \item \url{https://github.com/AlexaOrton/BH-SPD-CorrMat-Nets}
\end{enumerate}

\bibliographystyle{abbrvnat}
\bibliography{AOTG-SPDNET}

\end{document}